\documentclass[12pt,graphicx]{article}
\usepackage{hyperref}

\usepackage{graphicx}
\usepackage{amssymb}

\def\be{\begin{equation}}
\def\ee{\end{equation}}
\def\bea{\begin{eqnarray}}
\def\eea{\end{eqnarray}}

\def\IR{\relax{\rm I\kern-.18em R}}
\def\binomial#1#2{\left(\,{\buildrel
{\raise4pt\hbox{$\displaystyle{#1}$}}\over
{\raise-6pt\hbox{$\displaystyle{#2}$}}}\,\right)}

\def\[{\lfloor{\hskip 0.35pt}\!\!\!\lceil}
\def\]{\rfloor{\hskip 0.35pt}\!\!\!\rceil}

\def\a{\alpha}
\def\b{\beta}

\def\d{\delta}

\def\g{\gamma}

\def\m{\mu}

\def\s{\sigma}

\def\F{\Phi}

\def\J{\Psi}

\def\P{\Pi}

\def\S{\Sigma}

\def\cu{{\cal U}}

\newcommand{\drawsquare}[2]{\hbox{%
\rule{#2pt}{#1pt}\hskip-#2pt
\rule{#1pt}{#2pt}\hskip-#1pt
\rule[#1pt]{#1pt}{#2pt}}\rule[#1pt]{#2pt}{#2pt}\hskip-#2pt
\rule{#2pt}{#1pt}}

\renewcommand{\Box}{\,\raisebox{-.45pt}{\drawsquare{7}{0.6}}\,}

\def\bo{{\raise.15ex\hbox{\large$\Box$}}}
\def\pa{\partial}                                       

\def\TH{{\raise.2ex\hbox{$\displaystyle \bigodot$}\mskip-4.7mu \llap H
\;}}
\def\face{{\raise.2ex\hbox{$\displaystyle \bigodot$}\mskip-2.2mu \llap
{$\ddot
        \smile$}}}                                      

\def\Bar#1{\overline{#1}}                       
\def\leftrightarrowfill{$\mathsurround=0pt \mathord\leftarrow \mkern-6mu
        \cleaders\hbox{$\mkern-2mu \mathord- \mkern-2mu$}\hfill
        \mkern-6mu \mathord\rightarrow$}
\def\dvec#1{\vbox{\ialign{##\crcr
        \leftrightarrowfill\crcr\noalign{\kern-1pt\nointerlineskip}
        $\hfil\displaystyle{#1}\hfil$\crcr}}}           

\catcode`@=11
\def\un#1{\relax\ifmmode\@@underline#1\else
        $\@@underline{\hbox{#1}}$\relax\fi}
\catcode`@=12

\def\frac#1#2{{\textstyle{#1\over\vphantom2\smash{\raise.20ex
        \hbox{$\scriptstyle{#2}$}}}}}                   
\def\sfrac#1#2{{\vphantom1\smash{\lower.5ex\hbox{\small$#1$}}
\over
        \vphantom1\smash{\raise.4ex\hbox{\small$#2$}}}} 
\def\bfrac#1#2{{\vphantom1\smash{\lower.5ex\hbox{$#1$}}\over
        \vphantom1\smash{\raise.3ex\hbox{$#2$}}}}       
\def\afrac#1#2{{\vphantom1\smash{\lower.5ex\hbox{$#1$}}\over#2}}    

\newskip\humongous \humongous=0pt plus 1000pt minus 1000pt

\newif\ifdtup

  \def\pp{{\mathchoice
          {
              \kern 1pt%
              \raise 1pt
              \vbox{\hrule width5pt height0.4pt depth0pt
                    \kern -2pt
                    \hbox{\kern 2.3pt
                          \vrule width0.4pt height6pt depth0pt
                          }
                    \kern -2pt
                    \hrule width5pt height0.4pt depth0pt}%
                    \kern 1pt
           }
            {
              \kern 1pt%
              \raise 1pt
              \vbox{\hrule width4.3pt height0.4pt depth0pt
                    \kern -1.8pt
                    \hbox{\kern 1.95pt
                          \vrule width0.4pt height5.4pt depth0pt
                          }
                    \kern -1.8pt
                    \hrule width4.3pt height0.4pt depth0pt}%
                    \kern 1pt
            }
            {
              \kern 0.5pt%
              \raise 1pt
              \vbox{\hrule width4.0pt height0.3pt depth0pt
                    \kern -1.9pt  
                    \hbox{\kern 1.85pt
                          \vrule width0.3pt height5.7pt depth0pt
                          }
                    \kern -1.9pt
                    \hrule width4.0pt height0.3pt depth0pt}%
                    \kern 0.5pt
            }
            {
              \kern 0.5pt%
              \raise 1pt
              \vbox{\hrule width3.6pt height0.3pt depth0pt
                    \kern -1.5pt
                    \hbox{\kern 1.65pt
                          \vrule width0.3pt height4.5pt depth0pt
                          }
                    \kern -1.5pt
                    \hrule width3.6pt height0.3pt depth0pt}%
                    \kern 0.5pt
            }
        }}

  \def\mm{{\mathchoice
                       {
                             \kern 1pt
               \raise 1pt    \vbox{\hrule width5pt height0.4pt depth0pt
                                  \kern 2pt
                                  \hrule width5pt height0.4pt depth0pt}
                             \kern 1pt}
                       {
                            \kern 1pt
               \raise 1pt \vbox{\hrule width4.3pt height0.4pt depth0pt
                                  \kern 1.8pt
                                  \hrule width4.3pt height0.4pt depth0pt}
                             \kern 1pt}
                       {
                            \kern 0.5pt
               \raise 1pt
                            \vbox{\hrule width4.0pt height0.3pt depth0pt
                                  \kern 1.9pt
                                  \hrule width4.0pt height0.3pt depth0pt}
                            \kern 1pt}
                       {
                           \kern 0.5pt
             \raise 1pt  \vbox{\hrule width3.6pt height0.3pt depth0pt
                                  \kern 1.5pt
                                  \hrule width3.6pt height0.3pt depth0pt}
                           \kern 0.5pt}
                       }}

\def\ad{{\dot{\alpha}}}
\def\bd{{\dot{\beta}}}

\def\dslash{\not{\hbox{\kern-2pt $\partial$}}}
\def\Dslash{\not{\hbox{\kern-4pt $D$}}}
\def\pslash{\not{\hbox{\kern-2.3pt $p$}}}
 \newtoks\slashfraction
 \slashfraction={.13}
 \def\slash#1{\setbox0\hbox{$ #1 $}
 \setbox0\hbox to \the\slashfraction\wd0{\hss \box0}/\box0 }

\font\ro=cmsy10                          
\def\kcr{{\hbox{\ro \char'170}}}                
\def\ktl{{\hbox{\ro \char'170}}}        
\def\ktr{{\hbox{\ro \char'170}}}        
\def\kbl{{\hbox{\ro \char'170}}}        
\def\kbr{{\hbox{\ro \char'170}}}        

\def\plpl{\raise-2pt\hbox{$\raise3pt\hbox{$_+$}\hskip-6.67pt\raise0.0pt
\hbox{$^+$}\hskip 0.01pt$}}
\def\mimi{\raise-2pt\hbox{$\raise3pt\hbox{$_-$}\hskip-6.67pt\raise0.0pt
\hbox{$^-$}\hskip 0.01pt$}}

\parskip=\medskipamount
\thicklines

\def\border{                                            
        \setlength{\unitlength}{1mm}
        \newcount\xco
        \newcount\yco
        \xco=-21
        \yco=12
        \begin{picture}(140,0)
        \put(\xco,\yco){$\ktl$}
        \advance\yco by-1
        {\loop
        \put(\xco,\yco){$\kcr$}
        \advance\yco by-2
        \ifnum\yco>-240
        \repeat
        \put(\xco,\yco){$\kbl$}}
        \xco=158
        \yco=12
        \put(\xco,\yco){$\ktr$}
        \advance\yco by-1
        {\loop
        \put(\xco,\yco){$\kcr$}
        \advance\yco by-2
        \ifnum\yco>-240
        \repeat
        \put(\xco,\yco){$\kbr$}}
        \put(-20,13){\tiny **University of Maryland * Center for
String and
         Particle  Theory* Physics Department***University of
Maryland *Center
        for String and Particle  Theory** }
        \put(-20,-241.5){\tiny **University of Maryland * Center for
String and
         Particle  Theory* Physics Department***University of
Maryland *Center
        for String and Particle  Theory** }
        \end{picture}
        \par\vskip-8mm}
\def\headpic{                                           
        \indent
        \setlength{\unitlength}{.4mm}
        \thinlines
        \par
        \begin{picture}(29,16)
        \put(165,16){\line(1,0){4}}
        \put(170,16){\line(1,0){4}}
        \put(180,16){\line(1,0){4}}
        \put(175,0){\line(1,0){4}}
        \put(180,0){\line(1,0){4}}
        \put(185,0){\line(1,0){4}}
        \put(169,0){\line(0,1){16}}
        \put(170,0){\line(0,1){16}}
        \put(179,0){\line(0,1){16}}
        \put(180,0){\line(0,1){16}}
        \put(184,0){\line(0,1){16}}
        \put(185,0){\line(0,1){16}}
        \put(169,16){\oval(8,32)[bl]}
        \put(170,16){\oval(8,32)[br]}
        \put(179,0){\oval(8,32)[tl]}
        \put(185,0){\oval(8,32)[tr]}
        \end{picture}
        \par\vskip-6.5mm
        \thicklines}
\def\title#1#2#3#4{\border\headpic {\hbox to\hsize{#4 \hfill UMDEPP #3}}\par
        \begin{center} \vglue .5in {\large\bf #1}\\[.6in]
        {#2}\\[.1in] {\it Department of Physics and Astronomy}\\
        {\it University of Maryland, College Park, MD 20742}\\[1.5in]
        {\bf ABSTRACT}\\[.1in] \end{center} \begin{quotation}}
\def\Title#1#2#3#4#5#6#7{\border\headpic
        {\hbox to\hsize{#7 \hfill UMDEPP #6}}\par
        \begin{center} \vglue .4in {\large\bf #1}\\[.4in]
        {#2}\\[.1in] {\it Department of Physics and Astronomy}\\
        {\it University of Maryland, College Park, MD 20742}\\[.1in]
        {#3}\\[.1in] {\it {#4}}\\ {\it {#5}}\\[.5in] {\bf ABSTRACT}\\[.1in]
        \end{center} \begin{quotation}}                 
\def\endtitle{\end{quotation}\newpage}                  

\def\qd{{\kern0.5pt
                   q \kern-5.05pt \raise5.8pt\hbox{$\textstyle.$}\kern
0.5pt}}

\makeatletter

    \@addtoreset{equation}{section}%
\makeatother

\begin{document}
\def\gfrac#1#2{\frac {\scriptstyle{#1}}
         {\mbox{\raisebox{-.6ex}{$\scriptstyle{#2}$}}}}
\def\gg{{\hbox{\sc g}}}
\border\headpic
{\hbox to\hsize{January 2005 \hfill UMDEPP 05-038}}
{\hbox to\hsize{~\hfill UFIFT-HEP 05-4}}
\par
\setlength{\oddsidemargin}{0.3in}
\setlength{\evensidemargin}{-0.3in}
\begin{center}
\vglue .10in
{ \large\bf
Massive 4D, ${\cal N}=1$
Superspin 1 \& 3/2 Multiplets and \\
Dualities
\  }
\\[.25in]
I. L. Buchbinder$^\bullet$\footnote{joseph@tspu.edu.ru},
S. James Gates, Jr.$^\dag$\footnote{gatess@wam.umd.edu},
S. M.
Kuzenko\footnote{kuzenko@cyllene.uwa.edu.au}${}^\ast$,
and J.
Phillips$^{\dag\star}$\footnote{ferrigno@physics.umd.edu}\\
~\\
{\it {}$^\bullet$Department of Theoretical Physics, Tomsk State
Pedagogical
University\\ 634041 Tomsk, Russia}\\
~\\
{\it ${}^\dag$Center for String and Particle Theory\\
Department of Physics, University of Maryland\\
College Park, MD 20742-4111 USA}\\
~\\
{\it ${}^\ast$School of Physics, The University of Western Australia\\
Crawley, W.A. 6009, Australia}\\
~\\
{\it ${}^\star$Institute for Fundamental Theory\\
Department of Physics, University of Florida,\\
  Gainesville, FL, 32611, USA}\\
~\\
{\bf ABSTRACT}
\end{center}
\begin{quotation}
{Lagrangians for several new off-shell 4D, ${\cal N}=1$
supersymmetric descriptions of massive superspin-1 and
superspin-3/2 multiplets are described.  Taken together
with the models previously constructed, there are now
four off-shell formulations for the massive  gravitino
multiplet (superspin-1) and six off-shell formulations for
the massive graviton multiplet (superspin-3/2).  Duality
transformations are derived which relate some of  these
dynamical systems.}

${~~~}$ \newline
\endtitle

\section{Introduction}
\label{Introduction}
$~~\,~$ Different aspects of higher spin field theory
in various dimensions attract
considerable attention currently.
${}$First of all, higher spin fields and their
possible interactions bring about numerous
challenges for theoreticians.
More importantly, massive higher spin states
are known to be present in the spectra of the
string and superstring theories.
It is therefore quite natural to expect that,
in a field theory limit, the superstring theory
should reduce to a consistent
interacting supersymmetric theory of
  higher spin fields.

In four space-time dimensions,
Lagrangian formulations for massive fields
of arbitrary spin were constructed thirty years
ago \cite{SH}.  A few years later, the massive
construction of \cite{SH} was used to derive
Lagrangian formulations for gauge massless fields
of arbitrary spin \cite{Fronsdal}.
Since then, there have been published hundreds
of papers in which the results of \cite{SH,Fronsdal}
were generalized, (BRST) reformulated, extended,
quantized, and so forth.
Here it is  hardly possible to comment
upon these heroic follow-up activities.
We point out only several reviews \cite{REW} and
some recent papers \cite{DEV}.

One of the interesting directions in higher
spin field theory is the construction of
manifestly supersymmetric extensions
of the models given in \cite{SH,Fronsdal}.
In the massless case, the problem has actually
been solved in \cite{KSP,KS} (see \cite{BK}
for a review and \cite{KSG} for generalizations).
${}$For each superspin $Y>3/2$, these
publications provide two dually equivalent off-shell
realizations in  4D, ${\cal N }= 1$ superspace.
At the component level, each of the
two superspin-$Y$ actions
\cite{KSP,KS} reduces to a sum of the spin-$Y$
and  spin-$(Y+1/2)$ actions \cite{Fronsdal}
upon imposing a Wess-Zumino-type
gauge and eliminating the auxiliary fields.
On the mass shell, the only independent
gauge-invariant field strengths in these models
are  exactly the higher spin on-shell field strengths
first identified in ``Superspace''  \cite{GGRS}.
As concerns the massive case, off-shell higher spin
supermultiplets have never been constructed
in complete generality.

In 4D, ${\cal N}=1$ Poincar\'e
supersymmetry,
a massive multiplet of superspin $Y$
describes four propagating fields
with the same mass but different
spins $s$ $=$ $(Y-1/2, Y, Y, Y+1/2)$,
see, e.g.,  \cite{BK,GGRS} for reviews.
${}$The first attempts\footnote{Some preliminary results
were also obtained in \cite{BL}.}
to attack the problem of constructing
free off-shell massive higher spin supermultiplets
were undertaken in recent works
  \cite{BGLP1,BGLP2,GSS} that
  were concerned with deriving off-shell
realizations for the massive
{\it gravitino multiplet}  ($Y$ = 1)
and the massive {\it graviton multiplet}
($Y$ = 3/2). This led to two $Y$ = 3/2 formulations
constructed in \cite{BGLP1}
and one $Y$ = 1 formulation
derived in \cite{BGLP2}.
The results of \cite{BGLP1} were soon
generalized \cite{GSS} to produce
a third $Y$ = 3/2 formulation.

In the present letter, we continue the research
started in  \cite{BGLP1,BGLP2}
and derive two
new off-shell realizations for the massive gravitino
multiplet, and three new off-shell realizations
for the massive graviton multiplet.
Altogether, there now occur
four  massive $Y$ = 1 models
(in conjunction with the massive
$Y$ = 1 model constructed by
Ogievetsky and Sokatchev years ago \cite{OS2})
and six massive $Y$ = 3/2 models.
We further demonstrate that these realizations
are related to each other by duality transformations
similar to those which relate massive
tensor and vector multiplets, see \cite{K}
and references therein.

It is interesting to compare the massive
and massless results in the case of the $Y$ = 3/2
multiplet. In the massless case, there are three
building blocks to construct {\it {minimal}} \footnote{To
our knowledge, no investigations have occured for
the possible existence of
\newline $~~~~~~$ a massive
  {\it {non-minimal}}
$Y$ = 3/2 theory.}  linearized
supergravity \cite{GKP}. They correspond to
(i) old minimal supergravity
(see \cite{BK,GGRS} for reviews);
(ii) new minimal supergravity
(see \cite{BK,GGRS} for reviews);
(iii) the novel formulation derived in \cite{BGLP1}.
These off-shell $(3/2,2)$ supermultiplets,
which comprise all the supergravity
multiplets with $12+12$ degrees of freedom,
will be called  type I, type II and type III
supergravity multiplets\footnote{In the case
of type III supergravity, a nonlinear formulation
is still unknown.}
in what follows,
in order to avoid the use of unwieldy
terms like ``new new minimal''
or ``very new'' supergravity.
As is demonstrated below, each of the
massless type I---III
formulations admits a massive
extension, and the latter turns out to possess
a nontrivial  dual. As a result, we have now
demonstrated that there occur {\it {at}} {\it {least}}
six off-shell distinct massive $Y$ = 3/2
minimal realizations.

This paper is organized as follows.
In section 2 we derive two new
(dually equivalent) formulations
for the massive gravitino multiplet.
They turn out to be   massive extensions
of the two standard off-shell formulations
for the massless spin $(1,3/2)$ supermultiplet
discovered  respectively in \cite{FV,deWvH,GS}
and \cite{OS}.  In section 3 we derive three new
formulations for the massive graviton multiplet.
Duality transformations are also worked out
that  relate all the massive $Y$ = 3/2 models.
A brief summary of the results obtained is given
in section 4.  The paper is concluded by a technical
appendix.  Our superspace conventions mostly follow
\cite{BK} except the following two from
\cite{GGRS}: (i) the symmetrization of $n$ indices
does not involve a factor of $(1/n!) $;
(ii) given a four-vector
$v_a$, we define $v_{\un a} \equiv v_{\a \ad}
=(\s^a)_{\a \ad } v_a$.
\section{Massive Gravitino Multiplets}
\label{Massive Gravitino Multiplets}
$~~\,~$ We start by recalling the off-shell  formulation
for massless (matter) gravitino multiplet
introduced first in  \cite{FV,deWvH}
at the component level and then formulated
in \cite{GS} in terms of superfields
(see also \cite{GG}).
The action derived in \cite{GS} is
\bea
S_{(1,\frac 32)}[\J , V]
&=& \hat{S}[\J ]
+ \int d^8z\,\Big\{ \J^\a W_\a + {\Bar \J}_\ad
{\Bar W}^\ad \Big\}
- {1\over 4} \int d^6z \,W^\a W_\a~,
\label{tino}\\
&& \qquad \qquad
  W_\a = -{1 \over 4} \Bar D^2D^\a V~,
\nonumber
\eea
where
\bea
  \hat{S}[\J ]
&=& \int d^8z\, \Big\{
D^\a\Bar \J^{\dot\a}\Bar D_{\dot\a}\J_\a
-\frac 14\Bar D^{\dot\a}\J^\a\Bar D_{\dot\a}\J_\a
-\frac 14 D_\a \Bar \J_{\dot\a} D^\a \Bar \J^{\dot\a}
\Big\}~.
\label{S-hat}
\eea
This massless $Y$ = 1 model
is actually of some interest in the context of
higher spin field theory. As mentioned in
the introduction, there exist two dually
equivalent gauge superfield formulations
(called longitudinal and transverse)  \cite{KS}
for each massless {\it integer} $Y$
$\geq 1$, see \cite{BK} for a review.
The longitudinal series\footnote{The
transverse series terminates at a non-minimal
gauge formulation for the massless \newline
$~~~~~~$ gravitino multiplet
realized in terms of an unconstrained real scalar
$V$ and Majorana \newline
$~~~~~~$ $\g$-traceless
spin-vector ${\bf \J}_a=
(\J_{a \b}, \Bar \J_{a}{}^\bd)$,
with $\g^a {\bf  \J}_a=0$.}
terminates, at $Y =1$,  exactly at the action (\ref{tino}).

To describe a massive gravitino multiplet,
we introduce an action
$S=S_{(1,\frac 32)}[\J , V] + S_m[\J ,V]$,
where $S_m[\J ,V]$ stands for
the mass term
\bea
S_m[\J,V]=m \int d^8z\,\Big\{
\J^2
+\Bar \J^2
+\a m V^2
+ V\Big(\b D^\a \J_\a
+\b^* \Bar D_{\dot\a} \Bar \J^{\dot\a}\Big)
\Big\}~,
\label{S-m}
\eea
where $\a$ and $\b$ are  respectively real and complex
parameters.   These parameters should be fixed by the
requirement that the equations of motion be equivalent
to the constraints
\be
i \pa_{\un a} \Bar \J^\ad + m \J_\a = 0~, \qquad
D^\a \J_{\a}=0~, \qquad
\Bar D^2 \J_{\a}=0~~,~~
\label{mass-shell}
\ee
required to describe an irreducible on-shell multiplet
with $Y$ = $1$, see \cite{BK,BGLP2}.
In the space of spinor superfields
obeying the Klein-Gordon equation,
$(\Box -m^2) \J_\a =0$,
the second and third constraints
in (\ref{mass-shell}) are known to select
the $Y$ = 1 subspace \cite{BK}
(see also \cite{Sok}).
Without imposing
additional constraints
(such as the first one in \ref{mass-shell}),
the superfields $\J_\a $ and   $\Bar \J_\ad$
describe two massive $Y$ = 1
representations.  Generally, an irreducible representation
emerges if these superfields are also subject to
a reality condition of the form
\be
  \pa_{\un a} \Bar \J^\ad + m \,e^{i \varphi}\,
\J_\a = 0~, \qquad |e^{i \varphi} | =1~,
\ee
where $\varphi$ a constant real parameter.
As is obvious, the latter constraint implies
the Klein-Gordon equation. Applying a
phase transformation to $\J_\a$, allows
us to make the choice $e^{i \varphi} =-i$
corresponding to the Dirac equation.

The equations of motion
corresponding to
$S=S_{(1,\frac 32)}[\J , V] + S_m[\J ,V]$
are:
\bea
-\Bar D_{\dot\a}D_\a\Bar \J^{\dot\a}
+\frac 12\Bar D^2 \J_\a
+2m\J_\a
+W_\a
-\b m D_\a V&=&0~,
\label{y1newVeom} \\
\frac 12  D^\a W_\a
+\Big( \frac 14D^\a\Bar D^2 \J_\a
+\b mD^\a \J_\a
+c.c.\Big)
+2\a m^2V &=&0~.
\label{y1newHeom}
\eea
Multiplying (\ref{y1newVeom}) and
(\ref{y1newHeom})  by $\Bar D^2$ yields:
\bea
\Bar D^2\J_\a=
-2\b W_\a~, \qquad
  \Bar D^2D^\a \J_\a =-2{ \a \over \b }\, m\Bar D^2V~.
\eea
Next, substituting these relations into the contraction of $D^\a$
on (\ref{y1newVeom})  leads to:
\bea
mD^\a \J_\a = \frac 12 (\b+\b^*-1)D^\a W_\a
+\frac \b2\Big(1+{\a\over |\b|^2}\Big)mD^2V~.
\eea
Substitute these three results into (\ref{y1newHeom})
gives
\bea
\frac 12 (1-\b-\b^*)^2D^\a W_\a
+\frac 12m\Big(1+{\a\over |\b|^2}\Big)[\b^2D^2
+(\b^*)^2\Bar D^2]V
+2\a m^2 V =0~.
\eea
This equation implies that
$V$ is auxiliary, $V=0$, if
\be
\b+\b^*=1~, \qquad
\a=-|\b|^2~.
\label{conditions}
\ee
Then, the mass-shell conditions
(\ref{mass-shell})
also follow.\footnote{One can consider
more general action in which the term
$m ( \J^2 + \Bar \J^2) $
in (\ref{S-m})
is replaced  \newline
$~~~~\,~$ by
$ ( \m \, \J^2
+\m^*\,\Bar \J^2) $,
with $\m$ a complex mass parameter,
$|\m|=m$. Then, the first equation in  \newline
$~~~~\,~$
(\ref{conditions}) turns into
$\b/\m + (\b/\m)^* = 1/m$.}

The final action takes the form:
\bea
S[ \J,V] &=&
\hat{S}[\J ]
+ \int d^8z\,\Big\{ \J \, W
+ {\Bar \J} \,  {\Bar W} \Big\}
- {1\over 4} \int d^6z \,W^2
\label{ss1final}
\\
&~& ~~+~m \int d^8z\,\Big\{
\J^2
+\Bar \J^2
-|\b|^2 m V^2
+ V\Big(\b D \, \J
+\b^* \Bar D \, \Bar \J  \Big)
\Big\}~,
\nonumber
\eea
where $\b+\b^*=1$.
A superfield redefinition
of the form
$\J_\a\rightarrow \J_\a+ \d \, \Bar D^2\J_\a$
can be used to change some coefficients
in the action.

The Lagrangian constructed turns out
to possess a dual formulation. For simplicity,
we choose $\b =1/2$ in (\ref{ss1final}).
Let us consider, following \cite{K},
the ``first-order'' action
\bea
S_{Aux} &=& \hat{S}[\J ]
+  \int d^8z\,\Big\{ m(\J^2 + {\Bar \J}^2)
+\J \, W + {\Bar \J}\, {\Bar W}
-{m^2\over 4}  V^2
+ {m\over 2}  V( D\, \J
+ \Bar D\, \Bar \J ) \Big\}
\nonumber \\
&~& ~+~
{ 1 \over 2} \left\{
m \int d^6z \, \eta^\a \Big(W_\a
+ {1 \over 4} \Bar D^2D^\a V\Big)
- {1 \over 4} \int d^6z \,W^2 ~+~ c.c. \right\}~.
\label{aux1}
\eea
Here $W_\a$ and $\eta_\a$ are unconstrained chiral
spinor superfield, and there is no relationship
between $V$ and $W_\a$.
Varying $S_{Aux} $ with respect to $\eta_\a$
brings us back to (\ref{ss1final}).
On the other hand, if we vary
$S_{Aux} $ with respect to $V$
and $W_\a$ and eliminate these superfields,
we then  arrive at the following action:
\bea
\tilde{S} = \hat{S}[\J ]
&+&  \int d^8z\,\Big\{ m(\J^2 + {\Bar \J}^2)
+{1\over 4} \Big( D(\J+\eta) +
{\Bar D} ({\Bar \J} +  \Bar \eta ) \Big)^2
\Big\} \nonumber \\
&+& {1\over 8}
\left\{  \int d^6z \, \Big( 2m \eta
- \Bar D^2 \J \Big)^2
  ~+~ c.c. \right\}~.
\label{14}
\eea
Implementing here the shift
\be
\J_\a ~\to~ \J_\a - \eta_\a~~,~~
\ee
brings the action to the form
\bea
\tilde{S} = \hat{S}[\J ]
&+&{1\over 4} \int d^8z\,
\Big( D\,\J +
{\Bar D} \,{\Bar \J}  \Big)^2
- {1\over 2}  \int d^8z\,\Big\{ \J^\a \Bar D^2 \J_\a
+ \Bar \J_\ad D^2 \Bar \J^\ad \Big\}
\nonumber \\
&+& m \int d^8z\,\Big\{\J^2 + {\Bar \J}^2\Big\}
+ {m^2\over 2}
\left\{  \int d^6z \, \eta^2
  ~+~ c.c. \right\}~.
\eea
As is seen, the chiral spinor superfield $\eta_\a$
has completely decoupled! Therefore,
the dynamical system obtained is equivalent
to the following theory
\bea
S[\J] = \hat{S}[\J ]
&+&{1\over 4} \int d^8z\,
\Big( D\,\J +
{\Bar D} \,{\Bar \J}  \Big)^2
-{1\over 2}  \int d^8z\,\Big\{ \J^\a \Bar D^2 \J_\a
+ \Bar \J_\ad D^2 \Bar \J^\ad \Big\}
\nonumber \\
&+& m \int d^8z\,\Big\{\J^2 + {\Bar \J}^2\Big\}~~,~~
\label{mgravitino2}
\eea
formulated solely in terms of
the unconstrained spinor $\J_\a$
and it conjugate.
Applying the phase transformation
$\J_a \to i\, \J_\a$,  it is seen
that the  action obtained  is actually equivalent to
\bea
S[\J] = \hat{S}[\J ]
&-&{1\over 4} \int d^8z\,
\Big( D\,\J -
{\Bar D} \,{\Bar \J}  \Big)^2
+ m \int d^8z\,\Big\{\J^2 + {\Bar \J}^2\Big\}~.
\label{mgravitino3}
\eea

It is interesting to compare (\ref{mgravitino3})
with the action for massive $Y$ = 1 multiplet
obtained by Ogievetsky and Sokatchev \cite{OS2}.
Their model is also formulated solely
in terms of a spinor superfield. The
corresponding action\footnote{Setting
$m=0$ in (\ref{OS-action-2})
gives the model for massless gravitino
multiplet discovered in \cite{OS}.}
  is
\bea
  S_{OS} [\J]
=  \hat{S}[\J ]
+ {1\over 4}  \int d^8z\,
\Big(  D\J + {\Bar D} {\Bar \J}
\Big)^2
+ i \, m \int d^8z\,
\Big(  \J^2 - {\Bar \J}^2
\Big)~,
\label{OS-action-2}
\eea
see Appendix A for its derivation.\footnote{It
was argued in \cite{BGLP2}
that there are no Lagrangian formulations
for massive superspin-1 \newline
$~~~\,~~$ multiplet
solely in terms of an unconstrained
spinor superfield and its conjugate.
The \newline
$~~~\,~~$ ``proof'' given in \cite{BGLP2}
is incorrect, as shown by the  two
counter-examples (\ref{mgravitino3}) and
(\ref{OS-action-2}).}
The actions (\ref{mgravitino3}) and
(\ref{OS-action-2}) look similar,
although it does not seem possible
to transform one to the other off
the mass shell.

In fact, the model (\ref{14}),
which is equivalent to (\ref{mgravitino3}),
can be treated as a massive extension
of the Ogievetsky-Sokatchev model for
massless gravitino multiplet   \cite{OS}.
Indeed, implementing in (\ref{14}) the shift
\be
\J_\a ~\to \J_\a + {i \over 2m} \Bar D^2 \J_\a
~, \qquad
\eta_\a ~\to \eta_\a - {i \over 2m}
\Bar D^2 \J_\a~,
\ee
which leaves $\hat{S}[\J ] $ invariant,
we end up with
\bea
S[\J, \eta ]  =
S_{(1,\frac 32)}[\J , G]
&+&  m \int d^8z\,\Big\{ \J^2 + {\Bar \J}^2
+2(1 + i) \J \eta + 2(1 - i)\Bar \J  \Bar \eta \Big\}
\nonumber \\
&+& {m^2 \over 2} \left\{  \int d^6z \,  \eta^2
~+~ c.c. \right\}~,
\eea
where
\bea
S_{(1,\frac 32)}[\J , G]
=  \hat{S}[\J ]
&+&   \int d^8z\,
\Big( G +
{1\over 2} ( D\, \J + {\Bar D}\, {\Bar \J} )
\Big)^2  ~,  \\
G &=&{1\over 2} ( D^\a \eta _\a
+ \Bar D_\ad \Bar \eta^\ad )~.
\nonumber
\eea
Here $G$ is the linear superfield,
$D^2 G= \Bar D^2 G =0$, associated with
the chiral spinor $\eta_\a$ and its conjugate.
The action $S_{(1,\frac 32)}[\J , G] $ corresponds
to the Ogievetsky-Sokatchev formulation for
massless gravitino multiplet \cite{OS} as presented
in \cite{BK}.

Before concluding this section,
it is worth recalling one more possibility
to describe
the massless gravitino multiplet \cite{BK,GS}
\bea
S_{(1,\frac 32)}[\J , \F]
&=& \hat{S}[\J ]
-{1\over 2} \int d^8z\,\Big\{
\Bar \F \F  +
( \F +\Bar \F) ( D\, \J  +
\Bar D \, \Bar \J )
\Big\}~,
\label{g-fixed}
\eea
with $\F$ a chiral scalar, $\Bar D _\ad \F =0 $.
The actions (\ref{tino}) and (\ref{g-fixed})
can be shown to  correspond
to different partial gauge
fixings in  the mother theory
\bea
S_{(1,\frac 32)}[\J , V, \F]
= \hat{S}[\J ]
&+& \int d^8z\,\Big\{ \J \, W
+ {\Bar \J}  \, {\Bar W} \Big\}
- {1\over 4} \int d^6z \,W^2
\nonumber \\
&-&{1\over 2} \int d^8z\,\Big\{
\Bar \F \F  + (\F  +\Bar \F)
( D \,  \J  + \Bar D \, \Bar \J )
\Big\}~~,~~
\eea
possessing a huge gauge freedom,
see \cite{BK,GS} for more
details. The massive extension of (\ref{g-fixed})
was derived in \cite{BGLP2} and
the corresponding action is
\bea
S[\J , \F] = S_{(1,\frac 32)}[\J , \F]
+m\int d^8z\,(\J^2 + {\Bar \J}^2)
-{m\over 4}  \Big\{  \int d^6z \, \F^2
  + c.c. \Big\}~.
\eea
Unlike its massless limit,
this theory does not seem to admit
a nice dual formulation.

\section{Massive Graviton Multiplets}
\label{Massive Graviton Multiplets}

$~~\,~$ The massive $Y$ = 3/2 multiplet
(or massive graviton multiplet)
can be realized in terms of a  real (axial) vector
superfield $H_a$ obeying the  equations
\cite{BK,BGLP1,Sok}
\bea
\label{32irrepsp}
(\Box-m^2)H_a=0~, \quad
D^\a H_{\un a}=0~, \quad
\Bar D^{\dot\a} H_{\un a}=0 \quad
\longrightarrow  \quad \pa^{\un a}H_{\un a}=0~.
\eea
We are interested in classifying those supersymmetric
theories which generate these equations
as the equations of motion.

In what follows, we will use
a set of superprojectors \cite{SG}
for the real vector superfield $H_{\un a}$:
\bea
(\P^T_{1})H_{\un a}&:=&\frac 1{32}
\Box^{-2}\pa_{\dot\a}{}^\b
\{\Bar D^2,D^2\}\pa_{(\a}{}^{\dot\b}H_{\b)\dot\b}~, \\
(\P^T_{1/2})H_{\un a}&:=&
\frac 1{8\cdot3!}\Box^{-2}\pa_{\dot\a}{}^\b
D_{(\a}\Bar D^2D^\g
(\pa_{\b)}{}^{\dot\b}H_{\g\dot\b}
+\pa_{|\g|}{}^{\dot\b}H_{\b)\dot\b})~, \\
\label{trans}
(\P^T_{3/2})H_{\un a}&:=&
-\frac 1{8\cdot3!}\Box^{-2}\pa_{\dot\a}{}^\b
D^\g\Bar D^2D_{(\g}
\pa_{\a}{}^{\dot\b}H_{\b)\dot\b}~, \\
(\P^L_{0})H_{\un a}&:=&
-\frac 1{32}\pa_{\un a}
\Box^{-2}\{\Bar D^2,D^2\}
\pa^{\un c}H_{\un c}~, \\
\label{long}
(\P^L_{1/2})H_{\un a}&:=&
\frac 1{16}\pa_{\un a}\Box^{-2}D^\b\Bar D^2
D_\b\pa^{\un c}H_{\un c}~.
\eea
In terms of the superprojectors introduced,
we have \cite{GKP}
\bea
D^\g {\Bar D}^2 D_\g H_{\un a} &=&
-8\Box (  \P^L_{1/2} + \P^T_{1/2} +\P^T_{3/2})
H_{\un a} ~, \\
\pa_{\un a}\, \pa^{\un b} H_{\un b} &=&
-2 \Box ( \P^L_{0} +  \P^L_{1/2} ) H_{\un a}~,
\label{id2}\\
\left[D_\a , {\Bar D}_\ad \right]
\left[D_\b , {\Bar D}_\bd \right]
H^{\un b} &=& \Box (8  \P^L_{0}
- 24   \P^T_{1/2} ) H_{\un a}~.
\label{id3}
\eea
\subsection{Massive Extensions of Type I Supergravity}
\label{Massive Extensions of Type I Supergravity}
$~~\,~$ Consider the off-shell massive supergravity
multiplet  derived in \cite{GSS}
\bea
S^{({\rm IA})} [H, P] =
S^{({\rm I})} [H, \S]
- {1\over 2} m^2\int d^8z \,
\Big\{H^{\un a} H_{\un a}
-\frac 92 P^2 \Big\}~,
\label{IA}
\eea
where the massless part of the action takes the form
\bea
S^{({\rm I})} [H, \S] &=& \int d^8z \,
\Big\{
H^{\un a}\Box(
- \frac 13 \P^L_{0}+  \frac 12 \P^T_{3/2})H_{\un a}
-i(\S -\Bar \S ) \pa^{\un a}  H_{\un a}
- 3  \Bar \S  \S
\Big\}~, \\
&& \qquad
\S = -{1\over 4} \Bar D^2 P~, \qquad
\Bar P = P~~,~~
\nonumber
\eea
and this corresponds to a linearized form of type I (old minimal)
supergravity that has only appeared in the research literature
\cite{VariantSG}.  It has not been discussed in textbooks such as
  \cite{BK,GGRS}.  The distinctive feature {\it {unique}} to this theory
  is that its set of auxiliary fields contains one axial vector, one
scalar
  and one three-form ($S$, $C_{{\un a} \, {\un b} \, {\un c}} $,
  $A_{\un a}$).   Interestingly enough and to our knowledge,
  there has {\it {never}} been constructed a massive theory that contains
  the standard auxiliary fields of minimal supergravity ($S$, $P$,
  $A_{\un a}$).  This fact may be of some yet-to-be understood
  significance.

The theory with action $S^{({\rm IA})} [H, P]$
turns out to possess a dual formulation.
Let us introduce the ``first-order'' action
\bea
S_{Aux} &=&  \int d^8z \, \Big\{
H^{\un a}\Box( - \frac 13 \P^L_{0}
+  \frac 12 \P^T_{3/2})H_{\un a}
-\frac 12 m^2 H^{\un a}H_{\un a}  -U \pa^{\un a} H_{\un a}
\nonumber \\
&& ~~~~~~~~~~~
-\frac 32 U^2 + \frac 94 m^2 P^2
+3m V\Big( U + \frac14 \Bar D^2 P + \frac 14 D^2 P \Big)
\Big\}~,
\eea
where
$U$ and $V$ are real unconstrained superfields.
Varying $V$ brings us back to (\ref{IA}).
On the other hand, we can eliminate $U$ and
$P$ using their equations of motion.
With the aid of (\ref{id2}),
this gives
\bea
S^{({\rm IB})} [H, P] &=&  \int d^8z \, \Big\{
H^{\un a}\Box(  \frac 13 \P^L_{1/2}
+  \frac 12 \P^T_{3/2})H_{\un a}
-\frac 12 m^2 H^{\un a}H_{\un a}
\nonumber \\
&&~~~~~~~~~~- \frac{1}{16} V \{ \Bar D^2 , D^2 \} V
- m V  \pa^{\un a} H_{\un a}
+\frac 32 m^2 V^2 \Big\}~.
\label{IB}
\eea
This is one of the two formulations
for the massive $Y$ = 3/2 multiplet
constructed in \cite{BGLP1}.

\subsection{Massive Extensions of Type II Supergravity}
\label{Massive Extensions of Type II Supergravity}
$~~\,~$ Let us now turn to type II
(or new minimal) supergravity.
Its linearized action is
  \bea
S^{({\rm II})} [H, \cu] &=&\int d^8z\,
\Big\{H^{\un a}\Box(-\P^T_{1/2}+\frac 12\P^T_{3/2})H_{\un a}
+\frac 12\cu [D_\a,\Bar D_{\dot\a}]H^{\un a}
+\frac 32\cu^2\Big\}~,~~
\label{II}\\
&& \quad
\cu=D^\a \chi_\a+\Bar D_{\dot\a}\Bar \chi^{\dot\a}~,
\qquad \Bar D_\ad \chi_\a = 0~,
\nonumber
\eea
with $\chi_\a$ an unconstrained chiral spinor.
It possesses a unique massive extension
\bea
S^{({\rm IIA})} [H, \chi] &=&
S^{({\rm II})} [H, \cu]  - {1\over 2} m^2\int d^8z\,
H^{\un a}H_{\un a}
+3m^2 \left\{  \int d^6z \,  \chi^2
+c.c. \right\}
\label{IIA}
\eea
which is derived in Appendix B.

The theory (\ref{IIA}) admits a dual formulation.
Let us consider the following ``first-order'' action
\bea
S_{Aux} =\, \int d^8z\,
\Big\{H^{\un a}\Box(-\P^T_{1/2}+\frac 12\P^T_{3/2})H_{\un a}
-\frac 12 m^2 H^{\un a}H_{\un a}
+\frac 12\cu [D_\a,\Bar D_{\dot\a}]H^{\un a}
+\frac 32\cu^2
\nonumber\\
 ~~~~~~~~~~~- 6m V \Big( \cu - D^\a \chi_\a
-  \Bar D_{\dot\a}\Bar \chi^{\dot\a} \Big) \Big\}
+3m^2 \Big\{  \int d^6z \,  \chi^\a \chi_\a
+c.c. \Big\}~,~~~
\eea
in which $\cu$ and $V$ are real unconstrained
superfields. Varying $V$ gives the original action
(\ref{IIA}). On the other hand, we can eliminate
the independent real scalar $\cu$ and chiral spinor
$\chi_\a$ variables using their equations of motion.
With the aid of (\ref{id3}) this gives
\bea
S^{({\rm IIB})} [H, V] &=&\int d^8z\,
\Big\{H^{\un a}\Box(- \frac 13
\P^L_{0}+\frac 12\P^T_{3/2})H_{\un a}
-\frac 12 m^2 H^{\un a}H_{\un a}
\nonumber \\
&& \quad +mV  [D_\a,\Bar D_{\dot\a}]H^{\un a}
-6m^2 V^2\Big\}
-6  \int d^6z \,  W^\a W_\a~,
\label{IIB}
\eea
where $W_\a$
is the vector multiplet field strength
defined in (\ref{tino}).
The obtained  action (\ref{IIB}) constitutes
a new formulation
for massive supergravity multiplet.
\subsection{Massive Extensions of Type III Supergravity}
\label{Massive Extensions of Type III Supergravity}
$~~\,~$ Let us now turn to linearized type III
supergravity \cite{BGLP1}
  \bea
S^{({\rm III})} [H, \cu] &=&\int d^8z\,
\Big\{H^{\un a}\Box(\frac 13
\P^L_{1/2}+\frac 12\P^T_{3/2})H_{\un a}
+ \cu \pa_{\un a}  H^{\un a}
+\frac 32\cu^2\Big\}~,
\label{III}\\
&& \quad
\cu=D^\a \chi_\a+\Bar D_{\dot\a}\Bar \chi^{\dot\a}~,
\qquad \Bar D_\ad \chi_\a = 0~,
\nonumber
\eea
with $\chi_\a$ an unconstrained chiral spinor.
It possesses a unique massive extension
\bea
S^{({\rm IIIA})} [H, \chi] &=&
S^{({\rm III})} [H, \cu]  - {1\over 2} m^2\int d^8z\,
H^{\un a}H_{\un a}
-9m^2 \left\{  \int d^6z \,  \chi^2
+c.c. \right\}~,~~~
\label{IIIA}
\eea
and its derivation is very similar to that
of  (\ref{IIA}) given in Appendix B.

Similarly to the type II case considered earlier,
the theory (\ref{IIIA}) admits a dual formulation.
Let us introduce the ``first-order'' action
\bea
S_{Aux} &=& \int d^8z\,
\Big\{H^{\un a}\Box(\frac 13
\P^L_{1/2}+\frac 12\P^T_{3/2})H_{\un a}
-\frac 12 m^2 H^{\un a}H_{\un a}
+ \cu \pa_{\un a}  H^{\un a}
+\frac 32\cu^2
\nonumber \\
&& +3m V \Big( \cu - D^\a \chi_\a
-  \Bar D_{\dot\a}\Bar \chi^{\dot\a} \Big) \Big\}
-9m^2 \Big\{  \int d^6z \,  \chi^\a \chi_\a
+c.c. \Big\}~,
\eea
in which $\cu$ and $V$ are real unconstrained
superfields. Varying $V$ gives the original action
(\ref{IIIA}).
On the other hand, we can eliminate
the independent real scalar $\cu$ and chiral spinor
$\chi_\a$ variables using their equations of motion.
With the aid of (\ref{id2}) this gives
\bea
S^{({\rm IIIB})} [H, V] &=&\int d^8z\,
\Big\{H^{\un a}\Box(- \frac 13
\P^L_{0}+\frac 12\P^T_{3/2})H_{\un a}
-\frac 12 m^2 H^{\un a}H_{\un a}
\nonumber \\
&& \quad ~~~~~~~~-mV  \pa_{\un a} H^{\un a}
- \frac 32 m^2 V^2\Big\}
+{1\over 2}  \int d^6z \,  W^\a W_\a~,
\label{IIIB}
\eea
with the vector multiplet field strength
$W_\a$ defined  in eq. (\ref{tino}).
This is one of the two formulations
for the massive $Y$ = 3/2 multiplet
constructed in \cite{BGLP1}.  The other formulation is given
by the action (\ref{IB}).

\section{Summary}
\label{Summary}
$~~\,~$ We have formulated new free superfield
dynamical theories for massive multiplets of superspin
$Y$ = 1 and $Y$ = 3/2.  We have shown that these new theories
are  dually equivalent to the theories with corresponding superspin
given previously in the literature \cite{BGLP1,BGLP2,GSS}.
Although the theories with a fixed and specific value of $Y$
are on-shell equivalent, they differ from one another by distinctive
sets of auxiliary superfields (see discussion of this point in \cite{BGLP1}).
The existence of their varied and distinctive off-shell structures
together with their on-shell equivalence comes somewhat as
a surprise.

This surprise suggests that there is much remaining work to be done
in order to understand and classify the distinct off-shell representations
for all multiplets with higher values of $Y$
in both the massless and massive cases.
Our results raise many questions.   For example, for a fixed value
of $Y$ what massless off-shell representations possess massive
extensions?  How does the number of such duality related formulations
depend on the value of $Y$?
Are there even more off-shell possibilities
for the massless theories uncovered in the works of \cite{KSP,KS}?
Another obvious question relates to the results demonstrated
in the second work of \cite{KSG}.  In this past work, it was shown
that there is a natural way to combine 4D, $\cal N$ = 1 massless
higher spin supermultiplets into  4D, $\cal N$ = 2 massless
higher spin supermultiplets.   Therefore, we are led to expect that
it should be possible to  combine 4D, $\cal N$ = 1 massive
higher spin supermultiplets into  4D, $\cal N$ = 2 massive
higher spin supermultiplets.   As we presently only possess
{\it {four}} $Y$ = 1 and {\it {six}} $Y$ = 3/2 4D, $\cal N$ = 1
supermultiplets, the extension to 4D, $\cal N$ = 2 supersymmetry
promises to be an interesting study for the future.

All of these questions bring to the fore the need for
a comprehensive understanding of the role of duality
for arbitrary $Y$ supersymmetric
representations, of both the massless and massive varieties.
In  turn this raises the even more daunting specter
of understanding the role of duality within the context
of superstring/M-theory.  To our knowledge
the first time the question was raised about the possibility
of dually related superstrings was in 1985 \cite{GNi} and
there the question concerns on-shell dually related theories.
So for both on-shell and off-shell theories we lack
a complete understanding of duality.
The most successful descriptions of superstrings are of the
type pioneered by Berkovits (see \cite{BL} and references
therein).  As presently formulated, there is no sign of duality
in that formalism.  So does the superstring uniquely pick
out representations among the many
dual varieties suggested by our work?

\vskip.5cm
\noindent
{\bf Acknowledgments:}\\
The work of ILB was supported in part by the RFBR grant,
project No 03-02-16193, joint RFBR-DFG grant, project
No 02-02-04002, the DFG grant, project No 436 RUS 113/669,
the grant for LRSS, project No 125.2003.2 and INTAS grant,
project INTAS-03-51-6346.
The work of SJG and JP is
supported in part  by National Science Foundation
Grant PHY-0099544.
The work of SMK is supported in part by the
Australian Research Council.
\begin{appendix}
\section{Derivation of (\ref{OS-action-2})}
$~~\,~$ Let us start with the action
\bea
  S [\J]
=  \hat{S}[\J ]
+ {1\over 4}  \int d^8z\,
\Big(  D\, \J + {\Bar D} \,{\Bar \J}
\Big)^2
+ \int d^8z\,
\Big( \m \J^2 +  \m^* \, {\Bar \J}^2
\Big)~,
\label{OS-action}
\eea
where the functional $\hat{S}[\J]$ is
defined in (\ref{S-hat}), and
$\m$ is a complex mass parameter
to be specified later. The action
(\ref{OS-action}) with $\m=0$
describes the Ogievetsky-Sokatchev
model for the massless gravitino
multiplet \cite{OS}.
We are going to analyze whether this action
with $\m \neq 0$ can be used to consistently describe
the massive gravitino multiplet dynamics.
The equation of motion for $\J^\a$ is
\bea
-\Bar D_{\dot\a}D_\a\Bar \J^{\dot\a}
+\frac 12\Bar D^2 \J_\a
-\frac 12 D_\a ( D\, \J + {\Bar D}\, {\Bar \J}  )
+2\m\,\J_\a
&=&0~.
\label{OS-em1}
\eea
It implies
\bea
-\frac 14\Bar D^2  D_\a ( D\, \J + {\Bar D} \, {\Bar \J}  )
+\m \,\Bar D^2 \J_\a =0~,
\label{OS-em2}
\eea
and therefore
\bea
0&=& -\frac 14 D^\a\Bar D^2  D_\a
( D\, \J + {\Bar D} \, {\Bar \J}  )
+\m \,D^\a \Bar D^2 \J_\a
\label{OS-em3} \\
&=& -\frac 14 D^\a\Bar D^2  D_\a
( D\, \J + {\Bar D} \, {\Bar \J}  )
+\m \,\Bar D^2 ( D\, \J + {\Bar D}\, {\Bar \J}  )
+4i \m \,\pa^{\un a} \Bar D_\ad \J_\a ~.
\nonumber
\eea
Since the first term on the right is real and linear,
we further obtain
\bea
\m \,D^\a \Bar D^2 \J_\a &=&
  \m^* \, \Bar D_\ad D^2 \Bar \J^\ad~,
\label{OS-em4} \\
D^2 \Bar D^2 ( D\, \J + {\Bar D}\,  {\Bar \J}  )
&+& 4i \m \,\pa^{\un a}
D^2 \Bar D_\ad \J_\a =0~.
\label{OS-em5}
\eea
Since the operator $\Bar D^2 D^\a $
annihilates chiral superfields, applying
it to (\ref{OS-em1}) and making use of
(\ref{OS-em5}), we then obtain
\be
\Bar D^2 ( D\, \J + {\Bar D}\, {\Bar \J}  )
=D^2 ( D\, \J + {\Bar D} \, {\Bar \J}  )
  =0~.
\label{OS-em6}
\ee
Next, contracting $D^\a$ on (\ref{OS-em1})
and making use of
(\ref{OS-em6}) gives
\be
i \pa^{\un a}
( \Bar D_\ad \J_\a + D_\a \Bar \J_\ad)
+\m \, D\,\J =0~.
\label{OS-em7}
\ee
We also note that, due to (\ref{OS-em6}),
the equation (\ref{OS-em4}) is now
equivalent to
$ \pa^{\un a}
( \m \, \Bar D_\ad \J_\a -
  \m^* \, D_\a \Bar \J_\ad) =0$.
Therefore, with the choice
$ \m = i\, m$, where $m$ is real,
we end up with
\be
D \,\J = \Bar D \,\Bar \J =0~.
\label{OS-em8}
\ee
Then, eq.  (\ref{OS-em2}) becomes
\be
\Bar D^2 \J_\a = 0~.
\label{OS-em9}
\ee
Finally, the equation of motion
(\ref{OS-em1}) reduces to
\be
  \pa_{\un a} \Bar \J^\ad + m \,
\J_\a = 0~.
\label{OS-em10}
\ee
Eqs. (\ref{OS-em8}) -- (\ref{OS-em10})
define an irreducible $Y$ = 1 massive
representation.
They are equivalent to the equations
of motion in the Ogievetsky-Sokatchev model
(\ref{OS-action-2}).

\section{Derivation of (\ref{IIA})}
$~~\,~$ Let us consider an action
$S = S^{({\rm II})} [H, \cu]  + S_m [H, \chi]$,
where $S^{({\rm II})} [H, \cu] $ is the
type II supergravity action, eq. (\ref{II}),
and $S_m [H, \chi]$ stands for the mass term
\bea
S_m[H, \chi]=
-\frac 12m^2\int d^8z\, H^{\un a}H_{\un a}
+\frac 12\g m^2\int d^6z \,
  \chi^\a\chi_\a
+\frac 12\g^* m^2\int d^6\bar z \,
\bar \chi_{\dot\a}\bar\chi^{\dot\a}~,
\eea
with $\g$ a complex parameter.
The latter should be determined from
the requirement that
the equations of motion
\bea
\label{sweet}
\Box\Big[
\P_{3/2}^T
-2\P_{1/2}^T
\Big]H_{\un a}
-m^2H_{\un a}
+\frac 12[D_\a,\Bar D_{\dot\a}]\cu
&=&0~, \\
\label{tasty}
\frac 18\Bar D^2D_\a[D_\b,\Bar D_{\dot\b}]
H^{\un b}
+\frac 34\Bar D^2D_\a\cu
+m^2\g\chi_\a
&=&0~,~~
\eea
be equivalent to (\ref{32irrepsp}).
Since  $\cu$ is linear,
(\ref{sweet}) implies that $H_{\un a}$ is linear,
$D^2H_{\un a}=0$.
It is then possible to show
that $\Bar D^{\dot\a}H_{\un a}\propto \chi_\a$
on-shell.
To prove this proportionality,
first contract $\Bar D^{\dot\a}$ on (\ref{sweet})
and use the following identities:
\bea
\label{goodvibes}
\Bar D^2 D_\b [D_\a, \Bar D_{\dot\a}]H^{\un a}
&=&2i\Bar D^2 D^\a\pa_{(\a}{}^{\dot\a}H_{\b)\dot\a}~, \\
\Box\Bar D^{\dot\a}\P_{1/2}^T H_{\un a}&=&
-\frac i8\Bar D^2D^\d\pa_{(\a}{}^{\dot\b} H_{\d)\dot\b}
=-\frac 1{16}\Bar D^2 D_\b
[D_\a, \Bar D_{\dot\a}]H^{\un a}~,~
\eea
to arrive at:
\bea
+\frac 18\Bar D^2 D_\b [D_\a, \Bar D_{\dot\a}]H^{\un a}
+\frac 34\Bar D^2D_\a\cu
- m^2 \Bar D^{\dot\a}H_{\un a}
=0~.
\eea
Substituting the first two terms
with (\ref{tasty}) leads to:
\bea
\label{thebomb}
\g\chi_\a
+ \Bar D^{\dot\a}H_{\un a}=0~,
\eea
an upon substituting for $\cu$ in (\ref{tasty})
  by substituting (\ref{thebomb}) back in yields:
\bea
+\frac 18\Bar D^2D_\a[D_\b,\Bar D_{\dot\b}]H^{\un b}
-\frac 34\frac 1\g\Bar D^2D_\a[D^\b\Bar D^{\dot\b}
-\frac \g{\g^*}\Bar D^{\dot\b}D^\b]H_{\un b}
+m^2\g\chi_\a
=0~~.~~
\eea
This means that $\chi_\a$ will vanish if $\g$
is real and $\g=6$.  Equation (\ref{thebomb}) implies
that  $H_{\un a}$ is irreducible when $\chi_\a$ vanishes.
This means that $\P^T_{3/2}H_{\un a}=H_{\un a}$
and the Klein-Gordon equation
is obtained from (\ref{sweet}).
We therefore obtain (\ref{IIA}).

\end{appendix}


\end{document}